\documentclass[twocolumn,superscriptaddress,showpacs,preprintnumbers,amsmath,amssymb]{revtex4}

\newcommand{\bq}{\begin{equation}}
\newcommand{\ba}{\begin{eqnarray}}
\newcommand{\eq}{\end{equation}}
\newcommand{\ea}{\end{eqnarray}}

\newcommand {\tu} {{\tilde u}}
\newcommand {\tv} {{\tilde v}}
\newcommand {\tchi} {{\tilde \chi}}

\newcommand{\Ord}{\mathrm{O}}

\usepackage{amsfonts}
\usepackage{amssymb}
\usepackage{amsmath}
\usepackage{color}
\usepackage{graphicx}
\usepackage[pdftex,colorlin ks=true]{hyperref}
\usepackage{pdfpages}

\begin{document}

\title{Internal composite bound states in deterministic reaction diffusion models}
\author{Fred Cooper} 
\affiliation{Department of Earth and Planetary Sciences, Harvard University, Cambridge, MA 02138}
\affiliation{The Santa Fe Institute, 1399 Hyde Park Road, Santa Fe, NM 87501, USA}
\author{Gourab Ghoshal} 
\affiliation{Department of Earth and Planetary Sciences, Harvard University, Cambridge, MA 02138}
\author{Alec Pawling}  
\affiliation{Department of Earth and Planetary Sciences, Harvard University, Cambridge, MA 02138}
\author{Juan P\'erez-Mercader} \email{jperezmercader@fas.harvard.edu}
\affiliation{Department of Earth and Planetary Sciences, Harvard University, Cambridge, MA 02138}
\affiliation{The Santa Fe Institute, 1399 Hyde Park Road, Santa Fe, NM 87501, USA}



\begin{abstract}
 By identifying potential composite states that occur in the Sel'kov-Gray-Scott (GS) model, we show that it can be considered as an effective theory at large spatio-temporal  scales, arising from a more \textit{fundamental} theory (which treats these composite states as fundamental chemical species obeying the diffusion equation) relevant at shorter spatio-temporal scales.  When simulations in the latter model are performed as a function of a parameter $M = \lambda^{-1}$, the generated spatial patterns evolve at late times into those of the GS model at large $M$, implying that the composites follow their own unique dynamics at short scales. This separation of scales  is an example of  \textit{dynamical} decoupling in reaction diffusion systems. 
  \end{abstract}
\maketitle


Chemical and Biological Systems are in their details amongst the most complex systems we know. At larger spatio-temporal scales, however, when the microscopic details are not apparent, they display collective behaviors that manifest as relatively ÒsimpleÓ patterns. Some of these patterns show up in combustion or in biological systems as oscillatory patterns or even as self-replicating and evolving systems~\cite{Cross09,Grzybowski09}. Bridging the dynamics between the micro-- and macroscopic limits, remains a major challenge, especially acute in Biology.

In the late 60Õs, Sel'kov \cite{SelkovEJBC68} tried to model the chemical kinetics of the biochemistry of glycolysis by identifying and using 5 key chemical species out of the many known to be involved.  He noticed the existence in the problem of widely separated time scales and, exploiting this, he reduced the system to a meta-system with two ``proxy"  (fictitious) chemical species. Independently, years later, a particular case of Sel'kov's kinetics was discussed in the context of combustion and has since become known in the literature as the Gray-Scott (GS) \cite{GrayCES83} model. 
The GS model consists of two chemical species $U$ and $V$, with
corresponding chemical reactions:
\ba  \label{GS}
&&U+2V {\stackrel {\lambda}{\rightarrow}}~ 3 V, \nonumber \\
&&V {\stackrel {\mu}{\rightarrow}}~P, \quad U  {\stackrel {\nu}{\rightarrow}}~ Q, \nonumber \\
&&   {\stackrel {f}{\rightarrow}}~U.
\label{eq:GS}
\ea
There is a cubic autocatalytic step for $V$ at rate $\lambda$, and decay reactions at rates $\mu,\nu$ that transform $V$ and $U$ into inert products $P$ and $Q$. Finally $U$ is fed into the system at a rate  $f$. In the presence of diffusion the dynamics of the concentrations of the species is described by the equations,
\ba  \label{GS0} 
\frac{\partial v}{\partial t} && = \lambda u v ^2 - \mu v + D_v \nabla^2 v,  \nonumber \\
\frac{\partial u}{\partial t} && =f - \lambda u v^2 - \nu u+ D_u \nabla^2 u,
\label{eq:rdGS}
\ea 
where $u$ and $v$  (both functions of a $d+1$ dimensional space $(\vec x, t)$) denote the concentrations of $U,V$ while $D_u, D_v$ are the corresponding diffusion constants.   
(All parameters are necessarily positive in order to correspond to physically relevant quantities.)

The GS model with diffusion has been studied extensively in the literature, and despite its simple form has been found to contain rich spatio-temporal behavior (oscillations, chaos, etc.) as well as an extensive zoology of patterns (self-replicating spots, stripes, laces, spirals, etc.) that are phenomenologically interesting and suggestive~\cite{PearsonS93}. In fact these patterns, in principle, could be interpreted as displaying primitive forms of adaptation to the ``environment"  as the systems evolve in space and time~\cite{LesmesPRL03}. The form of the patterns depends on the relative values of the parameters describing the model at some ÒlargeÓ (pattern-size) scale where its collective behavior can be observed in computer simulations or in experiments~\cite{LeeN94}.

\begin{figure}[t!]
\centering 
\includegraphics[width=0.48\textwidth]{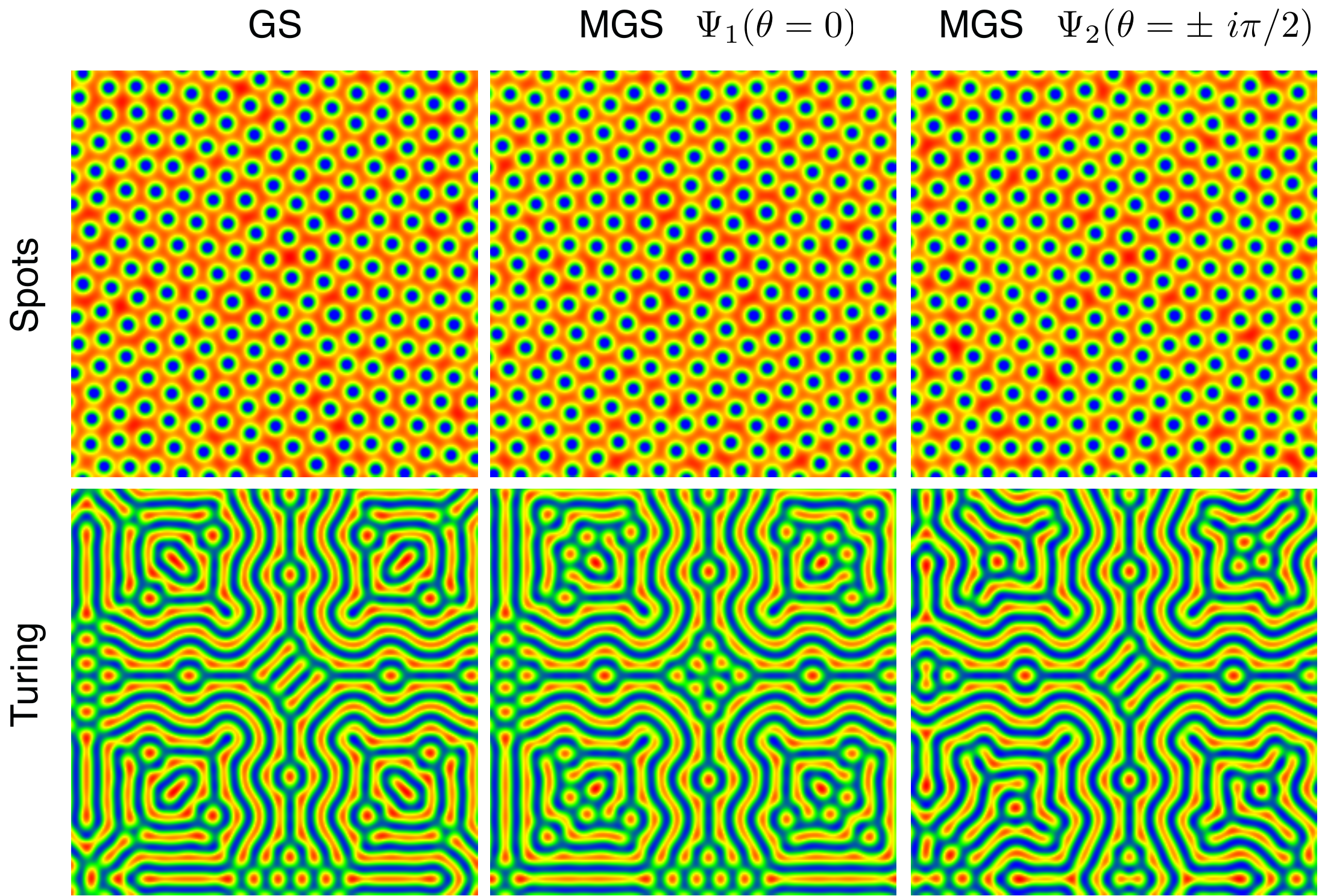}
\caption{The concentration of $U$ (red refers to high and blue, low concentrations) in the asymptotic state of the simulation of the original GS~\eqref{eq:rdGS} and modified MGS~\eqref{eq:mGS} models for two values of $\theta$ in the ``massive field" or low $\lambda$ limit ($\lambda = 5 \times 10^{-2})$. The simulations were carried out in a two-dimensional system of size $L_x = L_y = 200$, with periodic boundary conditions. The upper panel corresponds to the parameter regime that supports replication cascades ($\mu =0.75, \nu =0.15, f= 2.2, D_u/D_v = 5, D_{\Psi_{1,2}} = 1$) and the bottom panel shows the region supporting the formation of Turing patterns ($\mu =0.75,  \nu = 0.4, f = 4.2, D_u/D_v = 5, D_{\Psi_{1,2}} = 1$) in the GS model. In this limit, we see that the composite fields $\Psi_{1,2}$ ``decouple" from the dynamics of the system at late times, such that the patterns formed are qualitatively indistinguishable from the original model in multiple parameter regimes.}
\label{fig:fig111}
\end{figure}

In order to learn more about the properties of these models and thereby shed some light into the mechanisms leading to their phenomenology, it becomes imperative to \emph{fine-grain} the system in search of the properties of some putative \emph{internal} dynamics and degrees of freedom that collectively give rise to the observed larger scale behaviors. That is, to study the \emph{inverse} problem of what occurs when one tries to probe this system at spatio-temporal scales which are shorter than the ``effective" reaction-diffusion version of the GS model.  

As with all inverse problems, we are confronted with the fact that the internal structure of these models (i.e. their chemical pathways) may not be unique: there could, in principle, exist more than one short scale dynamics leading to the same observed larger scale behaviors.  At the same time there are many decompositions that \emph{do not}.
A way to eliminate the latter, is to enforce the criterion that the long-time behavior of the fine-grained system must coincide with that of the original large scale system. For those that do, there is the additional condition that,  in turn, the large scale system must \emph{not} feel the effects of the small scale dynamics.  In other words the dynamics at shorter scales must \emph{decouple}. A way to check whether this criterion is satisfied, is to compare the trajectories in phase space described by both systems, given arbitrary initial conditions.

If we select the initial conditions for the trajectories of the concentrations from a probability distribution function (see Supplementary Information Sec. S1), then one can write the collective evolution of the system by means of  a Path Integral \cite{Peliti, Doi, JC}.  When we do this for the GS model, the resulting Lagrangian turns out to be similar to that of the 
4-fermion theory of weak interactions \cite{Fermi}.  This analogy is striking since in the latter case there is the phenomenon of \emph{scale-separation}.  Indeed, at low energies and long distances, the theory involves the interactions of four fermions at a single space-time point. However at high energies and short distances, these local quartic interactions can be decomposed into a pair of cubic interactions each involving a boson (with large mass) and two fermions. The correspondence between the two limits (high and low energy)  is made by identifying the massive boson as  quadratic combinations of the fermionic  fields at low energies. Thus what looked like a quartic interaction at a coarse level actually arises from  a cubic interaction at a finer grain, where the composite of the fermions (the boson) is a real object (not just a mathematical artifact) obeying its own dynamics that effectively \emph{decouples} from the large scale dynamics~\cite{Weisberger,Galvan87}.
   
This field theory example suggests an approach to identify  the underlying degrees of freedom in the GS model.  Using this methodology, one can search for the presence of collective Òbound statesÓ (equivalent in chemistry to intermediate complexes in the mechanism of the reaction) whose properties  will be constrained by the law of mass action and by stoichiometry.  We first show that the GS model~\eqref{eq:rdGS} can be derived from a  Lagrangian~\cite{ref:Cherubini} with quartic  local interactions and then introduce composite bound states through constraint equations reducing it to an intermediate combination of cubic  interactions.
These interactions  represent two chemicals combining to form a short lived intermediate composite state. We then convert the composites to dynamical concentration fields (in analogy to what is done in the theory of weak interactions) and investigate whether the new model reproduces at large temporal scales the same spatial patterns as the original GS model. In doing so, we find that the new more detailed model preserves the GS dynamics at larger spatio-temporal scales while following its own unique dynamics at shorter scales.

The reaction diffusion equations for the GS model can be obtained from the following Lagrangian (cf. SI and \cite{ref:Cherubini} for details)
\ba  \label{gs1}
L &&= \int dx  \left(  \tv \left[  \frac{\partial v}{\partial t} -  D_v \nabla^2 v + \mu v - \lambda u v^2 \right ]  \right. \nonumber \\
&&+  \tu  \left.  \left[  \frac{\partial u}{\partial t} -  D_u \nabla^2 u - f +  \nu u + \lambda u v^2 \right ]  \right),
\label{eq:L1}
\ea
where $\tilde{v}, \tilde{u}$ are response fields.

Note that the Lagrangian in  Eq.~\eqref{eq:L1} contains quartic local interactions: $-\lambda \tilde{v} u v^2 ~\textrm{and} +\lambda \tilde{u} u v^2$.  Factoring out the response fields leaves us with the cubic terms $\pm \lambda u v^2$.  Our aim is to convert this into a quadratic term that preserves the stoichiometry of the original chemical reactions.  Note that
the cubic term originates from the autocatalytic step in Eq.~\eqref{eq:GS}, and assumes the simultaneous inelastic collision of two molecules of $V$ with one of $U$, which is highly unlikely.  More likely is the inelastic collision of just two molecules. The simplest choices of intermediate reactions that can be constructed from the cubic piece (without the introduction of new chemical species or fractional powers) are
\ba
V+V &{\stackrel {\lambda}{\rightarrow}}~[VV] \nonumber \\
U+V &{\stackrel {\lambda}{\rightarrow}}~[UV], 
\ea
that then react with one molecule of $U$ and $V$ respectively to complete the autocatalytic step. These are equivalent to having fast intermediate steps in the chemical reactions listed in Eq.~\eqref{eq:GS}. Thinking of the states $[VV], [UV]$  as collective variables that represent intermediate degrees of freedom and chemical pathways operating at shorter spatio-temporal scales than in~\eqref{eq:GS}, we introduce bilinear concentration fields $\Psi_1 = \lambda v^2$ and  
$\Psi_2=  \lambda u v$.

We are next faced with the challenge of determining the preferred chemical pathway. At the level of information in the GS system~\eqref{eq:GS} this is impossible to determine.  In general, which pathway is chosen, depends on the specifics and the details such as initial conditions under which the reaction takes place. To account for this,
 we write the autocatalytic term in Eq.~\eqref{eq:rdGS} as a linear combination of $\Psi_{1,2}$
\bq
\lambda u v^2 = u \Psi_1 \cosh^2 \theta   -  v \Psi_2 \sinh^2 \theta.
\eq
Here $\theta$ is a continuous mixing-angle, which takes into account any mixture of the two intermediate states, and serves as a proxy for simulating the specific conditions under which the chemical reactions take place. 

We then incorporate these terms  into~\eqref{eq:L1} via Lagrange multiplier fields $\tchi_{1,2}$ as

\bq
L_c = \int dx \frac{1}{\lambda}  \left[\tchi_1  ( \Psi_1 - \lambda v^2) + \tchi_2  ( \Psi_2 - \lambda u v)\right].
\eq

Recall from our earlier discussion, that in the 4-fermion model the composite field (the massive boson) was found to follow its own dynamics at shorter scales. Inspired by this, we allow the fields $\Psi_{1,2}$ to be dynamical fields obeying a reaction-diffusion equation and obtain the new Lagrangian
\ba  \label{gsnew5}
\tilde{L} &&= \int dx  \left(  \tv \bigg[  \frac{\partial v}{\partial t} -  D_v \nabla^2 v + \mu v -  u \Psi_1 \cosh^2 \theta + v \Psi_2 \right. \nonumber \\
&& \left. \times~\sinh^2 \theta \bigg] + \tu  \bigg[  \frac{\partial u}{\partial t} -  D_u \nabla^2 u  - f + \nu u  -  v \Psi_2 \bigg] \right. \nonumber \\
&& \left.  + ~\tchi_1    \bigg[  \frac{\partial  \Psi_1 } {\partial t} -  D_{\Psi_1}  \nabla^2  \Psi_1+ M \Psi_1 -     v^2 \bigg] \right. \nonumber \\
&&  \left.   + ~\tchi_2  \left[  \frac{\partial  \Psi_2 } {\partial t} -  D_{\Psi_2 } \nabla^2  \Psi_2 + M \Psi_2 -   u v  \right] \right).
\ea

Note that the fields $\Psi_{1,2}$ are created from the \emph{internal} dynamics, so there are no external sinks or sources for them. This is mathematically captured in the invariance of the $\Psi$ part of the Lagrangian under \emph{global} $U(1)$ rotations.  Assigning  $u, v$ a $U(1)$ charge of +1, implies that $\Psi_i$ must have charge +2 and $\tchi$ a charge of $-2$.  The last two terms in the Lagrangian Eq. (\ref{gsnew5}) are invariant under the $U(1)$ transformations,
\ba \label{u1}
&&u \rightarrow e^{i \alpha} u;~~ v \rightarrow e^{i \alpha} v; ~~ \tu ~\rightarrow e^{-i \alpha} \tu;~~   \tv ~\rightarrow e^{-i \alpha} \tv  \nonumber \\
&&\Psi_i \rightarrow e^{i 2 \alpha}   \Psi_i;~~ \tchi_i \rightarrow e^{-i 2 \alpha} \tchi_i,   
\ea
while the first two terms of  Eq. (\ref{gsnew5}) are {\em not} invariant \footnote{This symmetry will assume a greater importance when we consider the derivation of the GS model from a chemical master equation.}. 

\begin{figure}[t!]
\centering 
\includegraphics[width=0.48\textwidth]{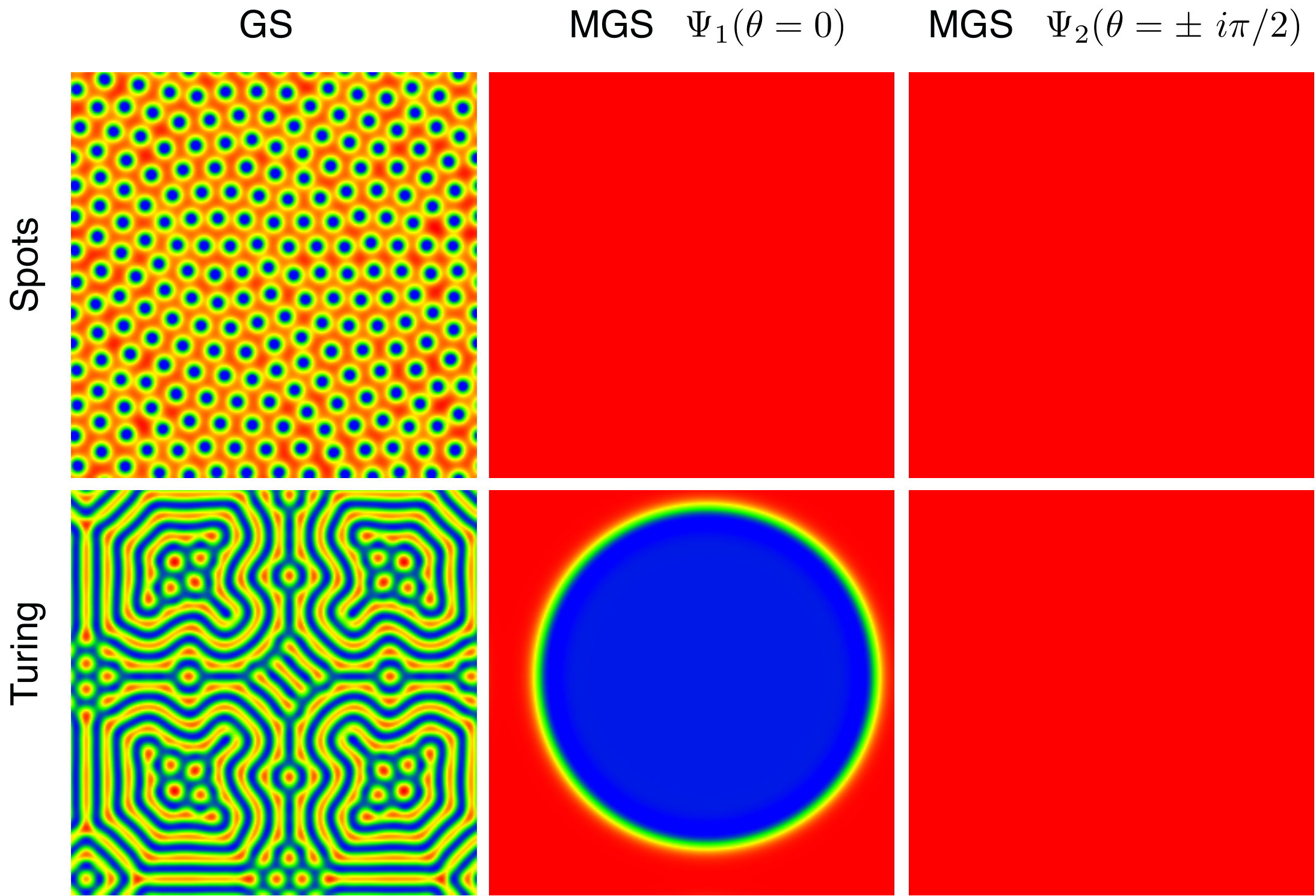}
\caption{Same as in Fig.~\ref{fig:fig111}, but now in the ``light field" or high $\lambda$ limit ($\lambda = 5.0$).  The parameters are the same as in Fig.~\ref{fig:fig111}, except $f$ which is set to $0.42$ and $0.44$ in the replication and Turing regimes respectively, as the phase regions governing what kinds of patterns form change with $\lambda$~\cite{MazinMCS96}.  We see that in this limit there is little correspondence between the GS  and MGS models and the set of equations in~\eqref{eq:mGS} correspond to a \emph{different} theory than the GS model.} 
\label{fig:fig222}
\end{figure}

Finally, by varying $\tilde{L}$ with respect to the response fields ($\tilde{u},\tilde{v},\tilde{\chi}_{1,2}$) we arrive at a modified set of Gray-Scott reaction diffusion equations (MGS), 
\ba
\frac{\partial v}{\partial t} && = u \Psi_1 \cosh^2 \theta - v \Psi_2 \sinh^2 \theta - \mu v + D_v \nabla^2 v,  \nonumber \\
\frac{\partial u}{\partial t} && =f  - u \Psi_1 \cosh^2 \theta +  v \Psi_2 \sinh^2 \theta - \nu u + D_u \nabla^2 u, \nonumber \\
\frac{\partial \Psi_1}{\partial t} &&= v^2 - M_1 \Psi_1 + D_{\Psi_1} \nabla^2 \Psi_1, \nonumber \\
\frac{\partial \Psi_2}{\partial t} &&= u v - M_2\Psi_2  + D_{\Psi_2} \nabla^2 \Psi_2,
\label{eq:mGS}
\ea 
which, by construction, have the same steady-state solutions as~\eqref{eq:GS}. 

It might seem that allowing the composites to be dynamical comes at the cost of four new parameters: $M_{1,2}$ and the two corresponding diffusion coefficients. However, we require the two models to be equivalent in the long time limit and this necessitates $M_{1,2} \gg 1$ so that in order to be consistent with the definition of the composite fields, $\Psi_{1,2} \rightarrow M_{1,2}^{-1} v^2$,
along with $M_{1,2} = \lambda^{-1}$.
Therefore the models are equivalent \emph{only} when the time scale at which the composite fields operate is much shorter than that at which pattern formation (in the original model) takes place. This separation of scales is governed by $\lambda$ and the equivalence between the two models occurs at the low $\lambda$ limit, i.e. $\lambda \ll 1$.  This limit implies that the autocatalytic step proceeds slowly \emph{relative} to the reactions for the composite fields. Thus in the ``frame of reference"  of the $\Psi$'s their formation takes place at a rate much faster than their reaction, allowing them to sample their own dynamics. Consequently varying $M$ \emph{independently} of $\lambda$ does \emph{not} correspond to the GS model.

The above can be tested by simulating the two models in relevant parameter regimes and checking whether they produce the same class of patterns at late times. 
The regimes in parameter space where various spatio-temporal patterns of the GS model are observed have been exhaustively mapped out~\cite{PearsonS93, MazinMCS96}.  In particular there are two static phases, the so-called \emph{red} phase where there is a non-zero concentration of $U$, $u_r = f / \nu$ with $v_r = 0$ and a \emph{blue} phase where both $u_b,v_b$ are non-zero.  The blue state has a Turing instability~\cite{Turing52} and is unstable with respect to spatially inhomogenous perturbations, leading to the formation of standing spatially periodic structures (Turing patterns).  The red state, on the other hand, is unstable to large amplitude localized perturbations resulting in the formation of a spot solution corresponding to enhanced concentrations of $V$ and depleted $U$.
Additionally, these spots are potentially unstable to replication and lead to a cascade that eventually fills the system~\cite{ReynoldsPRE97,DoelmanNl97,NishiuraPD99,MuratovJPAMG00}. Both these phenomena occur in different regions of the phase space of instabilities governed by the kinetic parameters $f,\lambda, \mu,\nu$ and the ratio of diffusion coefficients $D_u /D_v$.

In Fig.~\ref{fig:fig111} we plot the results of numerical simulations of both the GS and MGS models when $\lambda \ll 1$. The upper panel shows $u$ in the parameter regime of the GS model supporting spot replication, whereas the lower panel shows the regime where Turing patterns are formed.  The simulation confirms our qualitative argument regarding the correspondence between the two models in the long-time limit. The patterns in both parameter regimes are indistinguishable.  This clearly suggests that the composite fields $\Psi_{1,2}$ decouple from the dynamics of the system at late times in a manner such that pattern formation is unaffected. 
On the other hand when $\lambda$ is of $\Ord(1)$ the two models correspond to \emph{different} dynamics, a fact confirmed by Fig.~\ref{fig:fig222}, where we see that the MGS model generates neither replication cascades nor Turing patterns as occurs in this regime for the GS model.  

\begin{figure}[t!]
\centering 
\includegraphics[width=0.45\textwidth]{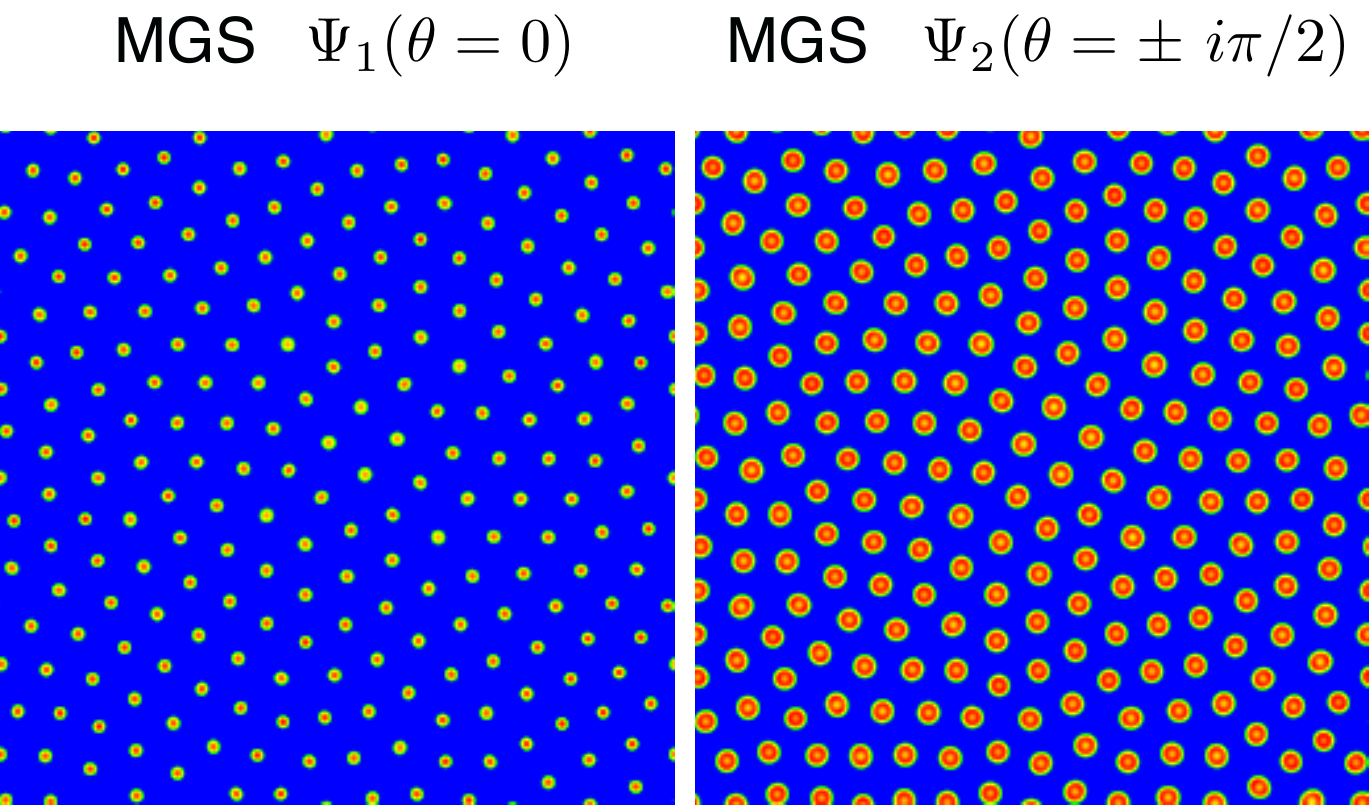}
\caption{The concentrations of $\Psi_1$ (left panel) and $\Psi_2$ (right panel) in the same simulation scheme as in Fig.~\ref{fig:fig111}. The concentration of $\Psi_1$ is co-located with that of $V$ whereas $\Psi_2$ has an annulus structure that is a combination of enhanced regions of $U,V$.}
\label{fig:fig333}
\end{figure}

In order to ensure the decoupling of the composites $\Psi_{1,2}$ we still need to study the evolution of their profiles in the $\lambda$ limit where the two models are equivalent.  When there is decoupling, one expects these profiles to remain confined within the longer scale effective--$U$ and --$V$ concentrations. 
In Fig.~\ref{fig:fig333} we show the concentrations of $\Psi_1$ (left panel) and $\Psi_2$ (right panel) for the same simulation scheme as in Fig.~\ref{fig:fig111}. The concentration profile of  $\Psi_1$ is co-located with $v$, in the sense that both have peak concentration values at  the \emph{same} spatial locations. On the other hand, $\Psi_2$ has an annulus structure consisting of a trough surrounded by a ring of enhanced concentration which is further surrounded by another depleted region. Taken together this suggests that the composite fields indeed act like bound states of the combinations $[VV]$ and $[UV]$, where the annulus structure in the latter case is a consequence of the fact that enhanced concentrations of $U$ correspond to depleted concentrations of $V$ and vice-versa.

In conclusion, we have studied the inverse problem of scale separation for the well-known example of the reaction diffusion Gray-Scott model. In some sense, the kinetic part of this problem is the reverse of Sel'kov's asymptotic time scale separation for the Krebs TCA Cycle~ \cite{SelkovEJBC68}, and is a first step to fine-grain such systems.  That is, to search for the internal structure that these systems should have in order for them to yield at the larger scales, the observed cooperative phenomenology.

To do this we have introduced the response field formalism~\cite{JC} and drawn on an analogy between the resulting Lagrangian and the 4-fermion theory of the weak interactions. This leads to an unique identification of composites which turn out to be dynamical intermediate states  of combinations of the chemical species $U$ and $V$. We see that this fine-graining procedure increases the details of the chemical dynamics, and where we initially had two chemical species operating at large spatio-temporal scales, we now have four at shorter scales: the two outer species $U,V$ and the internal composites $\Psi_{1,2}$ which are constrained by the stoichiometry of the original chemical reactions. Interestingly, we find that terms in the Lagrangian~\eqref{gsnew5} involving the composites are invariant to a global $U(1)$ symmetry. 
The scale at which the composites are active is determined by $M =\lambda^{-1}$.  We have confirmed by numerical simulations that the late time behaviors of both the effective $U,V$ and the full $U,V,\Psi_{1,2}$ models indeed show decoupling in the limit of $M \gg 1$. 

Finally, the methodology described here applies to systems more general than the GS model. Within the constraints of the relevant stoichiometry, a similar procedure of fine-graining can be applied to study \emph{any} chemical system with non-linear interaction terms.  This allows one to reconstruct hierarchies within the deterministic dynamics of a wide class of chemical systems.

\begin{acknowledgments}
This work was carried out with support from Repsol S. A.. The funders had no role in study design, data collection and analysis, decision to publish, or preparation of the manuscript.
\end{acknowledgments}

\end{document}